\documentclass[aps,prl,twocolumn,floatfix,showpacs,superscriptaddress,nofootinbib]{revtex4-2}
\usepackage{amsmath,amsfonts,amssymb,bm}
\usepackage{graphicx}
\usepackage{color}
\usepackage{subfigure}
\usepackage{multirow}
\usepackage{textcomp}
\usepackage{slashed}
\usepackage{bbm}
\usepackage[bottom]{footmisc}
\usepackage{parskip}
\definecolor{purple}{rgb}{0.5,0,0.5}
\definecolor{blue}{rgb}{0.0,0,1.0}
\usepackage[colorlinks=true, pdfstartview=FitV, linkcolor=purple, citecolor= purple, urlcolor=blue]{hyperref}

\newcommand{\beq}{\begin{equation}}
\newcommand{\eeq}{\end{equation}}
\newcommand{\ba}{\begin{array}}
\newcommand{\ea}{\end{array}}
\newcommand{\bea}{\begin{align}}
\newcommand{\eea}{\end{align}}
\newcommand{\bi}{\begin{itemize}}
\newcommand{\ei}{\end{itemize}}
\newcommand{\ben}{\begin{enumerate}}
\newcommand{\een}{\end{enumerate}}
\newcommand{\bc}{\begin{center}}
\newcommand{\ec}{\end{center}}
\newcommand{\bl}{\begin{flushleft}}
\newcommand{\el}{\end{flushleft}}
\newcommand{\br}{\begin{flushright}}
\newcommand{\er}{\end{flushright}}

\newcommand\comment[1]{ \hbox{[{\it Comment suppressed here.}\/]} }
\newcommand\hide[1]{}
\newcommand{\skipover}[1]{}

\begin{document}
\title{Empirical Determination of the Kaon Distribution Amplitude}

\author{Lei Chang}%
\affiliation{School of Physics, Nankai University, Tianjin 300071, China}
\author{Yu-Bin Liu}
\affiliation{School of Physics, Nankai University, Tianjin 300071, China}
\author{Kh\'epani Raya}
\affiliation{Department of Integrated Sciences and Center for Advanced Studies in Physics, Mathematics and Computation, University of Huelva, E-21071 Huelva, Spain.}
\author{M. Atif Sultan}%
\affiliation{School of Physics, Nankai University, Tianjin 300071, China}
\affiliation{Centre  For  High  Energy  Physics,  University  of  the  Punjab,  Lahore  (54590),  Pakistan}

\date{\today}

\begin{abstract}
We propose a data-driven approach to extract the Kaon leading-twist distribution amplitude (DA) from empirical information on the ratio of the neutral-to-charged kaon electromagnetic form factors, $\mathcal{R}_K$. Our study employs a two-parameter representation of the DA at $\zeta=2$ GeV, designed to capture the expected broadening and asymmetry of the distribution, as well as the soft endpoint behavior predicted by quantum chromodynamics (QCD). Our leading-order analysis of the latest experimental measurements of $\mathcal{R}_K$ reveals that the extracted DA exhibits a somewhat significant skewness, with the first symmetric moment approximately  $\langle 1-2x \rangle_K= 0.082(7)$. On the other hand, the brodaness and general shape of the produced distributions show a reasonable consistency with contemporaty lattice and continuum QCD analyses. These findings highlight the importance of accurately determining the profile of the DA, especially the skewness and its relation to $SU_F(3)$ flavor symmetry breaking, as well as the inclusion of higher-order effects in the hard-scattering kernels for analyzing data at experimentally accessible scales.
\end{abstract}

\maketitle

\noindent
\emph{Introduction.--}
The BESIII experiment has recently achieved a groundbreaking measurement of the ratio of neutral-to-charged kaon electromagnetic form factors (EMFFs) in the large-momentum-transfer regime $(12<Q^{2}<25$ GeV$^2)$,\,\cite{BESIII:2023zsk}. A constant value for this ratio $0.21\pm0.01$ was determined. Its small yet notable size exposes the $SU_F(3)$ flavor symmetry breaking (FSB) in the kaon wave function. Notably, the BESIII result aligns with a previous measurement at $|Q^2|=17.4\,\text{GeV}^2$\,\cite{Seth:2013eaa}. This exploration identifies an apparent discrepancy between the empirical extractions of the ratio of charged pion-to-kaon EMFFs and the predictions from quantum chromodynamics (QCD) at leading-order (LO) in perturbation theory,\,\cite{Farrar:1979aw,Lepage:1979zb,Lepage:1980fj,Efremov:1979qk}, where the so-called distribution amplitude (DA) plays a pivotal role. The results of\,\cite{Seth:2013eaa} are also seemingly inconsistent with projections based on non-perturbative frameworks\,\cite{Gao:2017mmp}. On its part, the recent analysis from Ref.\,\cite{Chen:2023byr} derives next-to-next-to-leading order (NNLO) corrections to the LO hard-scattering formulae (HSF) from Refs.\,\cite{Farrar:1979aw,Lepage:1979zb,Lepage:1980fj}, expanding the next-to-leading order (NLO) results from Refs.\,\cite{Field:1981wx,Dittes:1981aw,Sarmadi:1982yg,Braaten:1987yy}. Subsequently, employing recent lattice QCD (lQCD) determinations for the leading-twist kaon DA\,\cite{LatticeParton:2022zqc}, the exploration from Ref.\,\cite{Chen:2024oem} underscores the significant impact of higher-order contributions on the large-$Q^2$ regime of the EMFFs. Higher twist and other effects could also supplement the pQCD prediction\,\cite{Chai:2025xuz}. These observations is reinforced by the lQCD computation of the pion and kaon EMFFs at large momenta\,\cite{Ding:2024lfj}. Current and planned experimental efforts will certainly provide valuable insights on these aspects\,\cite{Accardi:2012qut,Accardi:2023chb}.

For the above, the importance of accurately determining the kaon DA is clear, but this need extends further. Kaons and pions are the Nambu-Goldstone bosons of dynamical chiral symmetry breaking (DCSB), so their existence and properties are deeply connected to the mass generation mechanisms in the Standard Model\,\cite{Roberts:2021nhw,Ding:2022ows,Raya:2024ejx}. In the absence of Higgs fields (HF) these states would be massless and identical. The structural differences observed in real life are arise from the $SU_F(3)$ FSB, whose size is controlled by the interplay between HF and QCD's mass generation. The shape of the DA is highly sensitive to this confluence: its broadness reflects the effects of DCSB, while its skewness would be influenced by both DCSB and the magnitude of FSB,\,\cite{Raya:2024ejx}. Several studies have demonstrated that the pion DA is broader than its asymptotic form at experimentally accessible energy scales  (\emph{e.g.} Refs.\,\cite{Chang:2013pq,Cui:2020tdf,Mikhailov:2016klg,Stefanis:2020rnd,Chang:2016ouf,Swarnkar:2015osa,Zhang:2020gaj,RQCD:2019osh,LatticeParton:2022zqc}). In contrast, lQCD and continuum analyses reveal that the heavy-quarkonia distributions are markedly narrower\,\cite{Blossier:2024wyx,Ding:2015rkn,Serna:2020txe,Xu:2025hjf}. Regarding the kaon, continuum Schwinger methods (CSMs)\,\cite{Shi:2014uwa,Cui:2020tdf}, holographic models\,\cite{Chang:2016ouf,Swarnkar:2015osa}, and lQCD\,\cite{Zhang:2020gaj,RQCD:2019osh,LatticeParton:2022zqc}, among others, all concur that its DA remains broad, though somewhat narrower than the pion's. Nonetheless, the degree of asymmetry is much less determined.

To address this gap, we propose an exploratory data-driven analysis for extracting the DA, which relies on leading-order HSF in QCD and the most recent experimental data on the neutral-to-charged kaon EMFFs\,\cite{BESIII:2023zsk}. This simplified LO approach permits:
\begin{itemize}
    \item Extract the kaon DA empirically, ultimately leading to a data-driven constraint on the flavor asymmetry within the kaon wave function.
    \item Evaluate the impact of skewness on the DA to determine how much this asymmetry alone can account for the measurements of the kaon EMFFs in the large-momentum transfer regime.
    \item Assess the validity of the current approach by comparing the resulting DAs with those from well-established frameworks.
    \item Establish the foundation for incorporating higher-order corrections. By first analyzing the role of the skewness in the DA at LO, we can better isolate and interpret the effects of higher-order corrections in future studies, enabling a more systematic exploration of the kaon structure.
\end{itemize}

The approach to be outlined not only addresses the key uncertainty in the pointwise behavior of the kaon DA, but also provides a clear pathway for reconciling theoretical predictions with experimental data. In the end, our work aims to advance the understanding of the kaon's internal structure, with a particular focus on the role of $SU_F(3)$ FSB in dictating its properties.

\noindent

\emph{Ratio of neutral-to-charged kaon form factors.--}  Through HSFs in QCD, the EMFF of a pseudoscalar meson $\textbf{P}$ can be expressed as a combination of hard and soft components. The former is, in principle, computable in perturbation theory; the soft part, on the other hand, encodes the non-perturbative effects via the DA\,\cite{Farrar:1979aw,Lepage:1979zb,Lepage:1980fj,Efremov:1979qk}. 

At leading-order, the EMFF would be completely determined by the contributions of the valence constituents, as revealed by the expression:
\begin{equation}
\label{eq:HSF}
Q^2 F_\textbf{P}(Q^2) \stackrel{Q^2 > Q_0^2}{\approx} 16 \pi \alpha_s(Q^2) f^2_\textbf{P} \sum_{f}e_{f} w^2_f(Q^2) \,,
\end{equation}
where $\alpha_s$ is one-loop strong running coupling and $f_\textbf{P}$ the meson's leptonic decay constant ($f_K\approx0.11$ GeV). The label $f$ indicates the flavor of the valence-quark, and $e_f$ its electric charge in terms of that of the fundamental one. Note the HSF hold at sufficiently high energies $Q_0^2 \gg \Lambda_{\text{QCD}}^2$, though the precise $Q_0^2$ value is not inherently determined from QCD principles. The weight-factor $w_{f}(Q^2)$ is linked to the soft part in the HSF. This is defined as follows:
\begin{equation}
\label{eq:Weight}
    w_{f}(Q^2)=\frac{1}{3}\int dx \,\frac{\varphi_\textbf{P}^f(x;Q^2)}{1-x}\,.
\end{equation}
Here, $\varphi_\textbf{P}^f(x;Q^2)$ represents the meson's leading-twist DA. Intuitively, it describes the likelihood of finding a valence-quark $f$ to carrying a momentum fraction $x$ of the meson's total momentum. Our choice of $Q^2$ as the defining scale of the DA indicates that this has also been adopted as the factorization scale in Eq.\,\eqref{eq:HSF}. Note also that the appearance of $\alpha_s$ and the DA in the HSF expose the scaling violations of QCD\,\cite{Raya:2024ejx,Yao:2024drm}.

\par
We assume isospin symmetry under which the up ($u$) and down ($d$) quarks are treated as identical except for their electric charges. For notational convenience, these quarks will be referred to as $l$-quarks. This symmetry allows us to express the ratio of the neutral-to-charged kaon form factor, $\mathcal{R}_K(Q^2)$, as:
\begin{equation}
\label{eq:RatioDef}
    \mathcal{R}_K(Q^2):=\frac{|F_{K^0}(Q^2)|}{|F_{K^+}(Q^2)|}\overset{Q^2>Q_0^2}{\approx}\left|\frac{-\frac{1}{3} w_{l}^{2}(Q^2)+\frac{1}{3}w_{s}^{2}(Q^2)}{\frac{2}{3} w_{l}^{2}(Q^2)+\frac{1}{3}w_{s}^{2}(Q^2)}\right|\,.
\end{equation}
Such quantity provides a measure of the relative contributions of the valence-quarks to the kaon form factors, reflecting not only the interplay between the electric charges but also, through the corresponding DAs, the effects of FSB induced by the mass generation mechanisms.
\par
The intuition above is further reinforced by considering the domain of asymptotically large energies ($Q^2\to\infty$), where the DAs converge to\,\cite{Farrar:1979aw,Lepage:1979zb,Lepage:1980fj,Efremov:1979qk}:
\begin{equation}
\label{eq:asymDA}
    \varphi_{\textbf{P}}^f(x;Q^2\to\infty)\to\varphi^{asy}(x)=6x(1-x)\,.
\end{equation}
Within this region, the weight factors $w_{l}$ and $w_s$ approach unity~\cite{Farrar:1979aw,lepage1980exclusive}, leading to a vanishing ratio $\mathcal{R}_K(Q^2\to\infty) \to 0$. This would be the same case, in the whole energy range, if $w_u$ and $w_s$ were identical. A non-zero value of $\mathcal{R}_K(Q^2)$ at finite values of $Q^2$ is therefore a reflection of the flavor asymmetry, measured through the asymmetry in the corresponding DAs. The skewness in the kaon DA encodes the differences in the momentum distributions of the $l$ and $s$-quarks within the kaon, pointing out the role of flavor-dependent effects in the electromagnetic structure of mesons.

\noindent
\emph{Distribution amplitude at 2 GeV.--} While various perspectives confirm certain features of the kaon DA, such as a broadness comparable to that of the pion (which itself is wider than the asymptotic profile $\varphi^{asy}(x)$ at accessible energies), the precise form of the kaon DA remains an open question. In particular, as previously noted, the extent of its asymmetry deserves special attention.
\par
For the purpose of our discussion, we consider a leading-twist DA defined at a resolution scale $\zeta_2=2$ GeV, expressed via a two-parameter functional form as follows:
\begin{equation}\label{eq:PDA1}
\varphi_\textbf{P}^l(x;\zeta_2)=\mathcal{N}_\textbf{P}\, \text{ln}\Big( 1+\frac{x(1-x)}{\rho_0^\textbf{P}+\rho_1^\textbf{P}(2x-1)} \Big)  \,;
\end{equation}
here $\rho_0^\textbf{P}$ and $\rho_1^\textbf{P}$ are the parameters to be determined, and $\mathcal{N}^\textbf{P}$ ensures the unit-normalization of the DA. Adopting this representation has important advantages. Firstly, the distribution's Mellin moments, 
\begin{equation}
  \langle (1-2x)^m \rangle^\textbf{P}_{\zeta_2}=\int_0^1 dx\, \varphi_\textbf{P}^l(x;\zeta_2)\, (1-2x)^m\,,  
\end{equation}
can be obtained algebraically. Secondly, despite the reduced number of parameters, the proposed form is capable of capturing the expected broadness and skewness of the light-meson DAs. For instance, the distributions reported in Ref.\,\cite{Cui:2020tdf}, obtained using sophisticated kernels (DB) within the CSMs framework, are accurately reproduced with:
\begin{equation}
\label{eq:paramsCSMDB}
    \rho_0^\pi=0.032\,,\rho_1^\pi=0\;;\,\,\rho_0^K=0.054\,,\rho_1^K=0.013\,.
\end{equation}
Meanwhile, the \emph{rainbow-ladder} (RL) expectations from Refs.\,\cite{Chang:2013pq,Shi:2014uwa}, are satisfied provided:
\begin{equation}
\label{eq:paramsCSMRL}
    \rho_0^\pi=0.003\,,\rho_1^\pi=0\;;\,\,\rho_0^K=0.133\,,\rho_1^K=0.082\,.
\end{equation}
Throughout the rest of the text, we will use these representations for the DB and RL expectations. Evidently, the asymmetry in $\varphi_\textbf{P}^l(x;\zeta_2)$ arises from $\rho_1^\textbf{P}$, whereas the broadness and all moments $\langle\xi^m\rangle^\textbf{P}_{\zeta_2}$ are influenced by both $\rho_0^\textbf{P}$ and $\rho_1^\textbf{P}$. This contrasts with the logarithmic representation proposed in Ref.\,\cite{Raya:2024ejx}, where even moments are independent of the parameter controlling the skewness. We choose to keep this correlation between the model parameters. Finally, a desirable characteristic of Eq.\,\eqref{eq:PDA1} is that it faithfully captures the endpoint behavior prescribed by QCD; that is, $\varphi_\textbf{P}^l(x\to1)\sim(1-x)$. 
\par
These features make our parametrization more advantageous compared to alternative forms. For instance, approaches based on a $C_j^{3/2}$-Gegenbauer polynomial expansion may require a large number of terms to achieve the desired accuracy\,\cite{Chang:2013pq}. Alternatively, representations employing a multiplicative factor $x^{\alpha}(1-x)^{\beta}$ yield an incorrect endpoint behavior. This, in turn, results in a poor estimation of the $\langle x^{-1},(1-x)^{-1}\rangle$ moments entering the HSF, preventing an accurate description of the large-$Q^2$ behavior of the EMFFs. By considering the form in Eq.\,\eqref{eq:PDA1}, we aim to provide a more physically motivated yet flexible representation of the kaon DA, which maintains consistency with QCD predictions and existing phenomenological observations.
\par
\begin{figure}[hptb!]
\includegraphics[width=8.6cm]{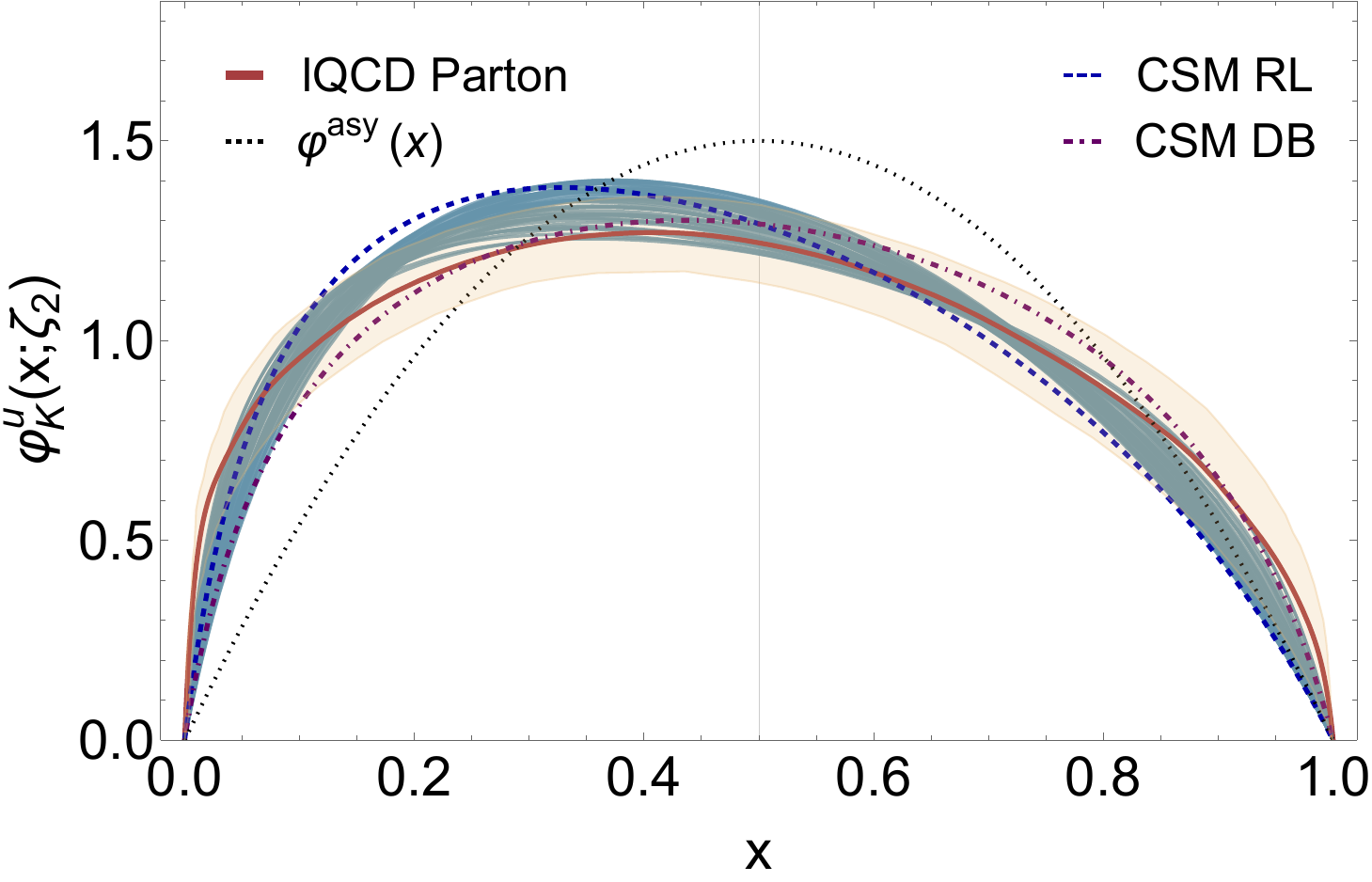}
\caption{Kaon leading-twist DAs at $\zeta=2\,\text{GeV}$, $\varphi_K^l(x;\zeta_2)$. Our set of replicas is drawn as light-blue solid lines. For comparison, we include results from lQCD~\cite{LatticeParton:2022zqc} and CSMs\,\cite{Shi:2014uwa,Cui:2020tdf}. The black-dotted curve represents the asymptotic profile $ \varphi^{\text{asy}}(x) = 6x(1-x)$. Corresponding low-order Mellin moments are provided in Table\,\ref{tab:momsDA}.
\label{fig:PDA}}
\end{figure}

In order to determine the model parameters that define $\varphi_K^u(x;\zeta_2)$, we adopt the following strategy:
\begin{enumerate}
    \item A random value $\rho_0^{(i)}$ is chosen from the interval $(0,0.2)$.
    \item Given $\rho_0^{(i)}$, the parameter $\rho_1^{(i)}$ is randomly selected within a range that ensures:
    \begin{equation}
        \langle \xi^2\rangle^{(i)} \in (0.22,0.26)\,.
    \end{equation}
    This constraint is informed by various continuum and lQCD studies\,\cite{Xu:2025hjf,Serna:2020txe,Shi:2014uwa,Cui:2020tdf,RQCD:2019osh,LatticeParton:2022zqc}.
    \item Once $\varphi_{(i)}^{I}$ is fully defined by the two parameters $\rho_{0,1}^{(i)}$, Eq.\,\eqref{eq:RatioDef} is applied to compute $\mathcal{R}^{(i)}_K(Q^2)$ over a discrete set of $Q^2_j$ values matching the experimental data points. 
    
    Note that for each $Q_j^2$, the DA is evolved from the starting scale $Q^2=\zeta_2^2$ to $Q_j^2$, according to the LO evolution equations\,\cite{Lepage:1979zb,Lepage:1980fj,Efremov:1979qk}. For this purpose, we set $\Lambda_{\text{QCD}}=0.234\,\text{GeV}$ and $n_f=4$ flavors.

    \item The resulting $\mathcal{R}_K^{(i)}(Q^2)$ is compared with the experimental data, where the errors bars are symmetrized. If the computed $\chi^2/d.o.f<2$,  the  $\{\rho_0^{(i)},\rho_1^{(i)}\}$ pair is retained.
    \item This process is repeated until 50 valid duplets are produced.
\end{enumerate}
The outcome of this procedure is shown in Fig.\,\ref{fig:PDA}. We observe that the generated set of DAs exhibits a notable agreement with the lQCD determination,\,\cite{LatticeParton:2022zqc}, though our results show a greater degree of asymmetry. Other evaluations also fall within the ballpark, \emph{e.g.}\,\cite{RQCD:2019osh,Chang:2016ouf,Swarnkar:2015osa}. The same figure offers a comparison with the results from CSMs, employing both RL and DB kernels\,\cite{Shi:2014uwa,Cui:2020tdf}. The RL profile exhibits a more pronounced skewness, effectively setting a boundary. In contrast, the kaon DA derived from the DB kernel features a more symmetric shape. The peak of the DA provides a way to quantify its skewness. In our case, the distribution reaches its maximum at 
\begin{equation}
    \label{eq:max}
    x^{\text{max}}=0.37(1)\;,
\end{equation}
reflecting a $25\%$ deviation from the symmetric limit $x^{\text{max}}=0.5$.  The typical trend for this shift is around $\sim20\%$\,\cite{LatticeParton:2022zqc,Cui:2020tdf,RQCD:2019osh}. Such a displacement is of the same order of the $f_K-f_\pi$ difference, quantities that serve as indicators of the strength of DCSB. This suggest that the skewness in the DA is primarily driven by QCD non-perturbative phenomena (mass generation) rather than the large disparity among the current-quark masses ($m_s/m_l\sim 20$)\,\cite{Raya:2024ejx,Ding:2022ows,Roberts:2021nhw}. Notable exceptions to this pattern include the RL result, where the peak of the distribution is deviated around $33\%$ from the symmetric case,\,\cite{Shi:2014uwa}, and the lQCD analysis from\,\cite{Zhang:2020gaj}, which, within uncertainties, yields a nearly symmetric kaon DA that resembles the asymptotic profile.

The evaluation carried out here leads to the following mean values:
\begin{eqnarray}
\label{eq:finalMoms}
    \langle \xi \rangle_{\zeta_2}^K=0.082(7)\,,\,\langle \xi^2 \rangle_{\zeta_2}^K=0.239(9)\,.
\end{eqnarray}
A comparison with the moments presented in Table\,\ref{tab:momsDA} suggests that our prediction for $\langle \xi \rangle_{\zeta_2}^K$ exceeds the general trend, being surpassed only by the RL result. This quantity also encapsulates the degree of skewness and, consequently, the strength of the $SU_F(3)$ FSB. Unsurprisingly, the kaon DA's dilation --  the effects of DCSB -- are accurately captured, as demonstrated by the precise agreement of  $\langle \xi^2 \rangle_{\zeta_2}^K$ with phenomenological expectations.

The inverse moments of the DA provide another measure of its asymmetry. Here we find ($\bar{x}=1-x$):
\begin{eqnarray}
\label{eq:finalMoms2}
    \,\langle 1/x \rangle_{\zeta_2}^K=4.03(18)\,,\,\langle 1/\bar{x} \rangle_{\zeta_2}^K=2.89(11)\,.
\end{eqnarray}
As revealed in Table\,\ref{tab:momsDA}, distributions with more asymmetric profiles (hence larger $\langle \xi \rangle_{\zeta_2}^K$ moments) lead to a greater difference between the inverse moments. Thus, these values also encode the degree of $SU_F(3)$ FSB. In this sense, and due to their relationship with HSFs, large virtuality EMFFs are extremely useful in their determination.

\begin{table}[h!]
\begin{tabular}[t]{l||c|c|c|c}
\hline
& $\langle \xi \rangle_{\zeta_2}^K$ & $\langle \xi^2 \rangle_{\zeta_2}^K$ & $\langle 1/x \rangle_{\zeta_2}^K$ & $\langle 1/\bar{x} \rangle_{\zeta_2}^K$\\
\hline
This work &  $\,0.082(7)\,$ & $\,0.239(9)\,$ & $\,4.03(18)\,$ & $\,2.89(11)\,$ \\
lQCD 22'\,\cite{LatticeParton:2022zqc}&  $\,0.065(31)^*\,$ & $\,0.258(32)\,$ & $\,4.18(23)^*\,$ & $\,3.28(23)^*\,$ \\
lQCD 20'\,\cite{Zhang:2020gaj}&  $\,0.002(69)^\dagger\,$ & $\,0.198(16)\,$ & $2.97(42)^\dagger$ & $2.94(42)^\dagger$ \\
lQCD 19'\,\cite{RQCD:2019osh} &  $\,0.032(12)\,$ & $\,0.231(4)\,$ & $\,3.43(8)^*\,$ & $\,3.28(23)^*\,$ \\
CSM DB\,\cite{Cui:2020tdf} &  $\,0.035(5)\,$ & $\,0.24(1)\,$ & $\,3.271\,$ & $\,3.21\,$ \\
CSM RL\,\cite{Shi:2014uwa}&  $\,0.11\,$ & $\,0.23\,$ & $\,4.20\,$ & $\,2.72\,$ \\
\hline
\end{tabular}
\caption{Low-order Mellin moments of the kaon DA $\varphi_K^l(x;\zeta_2)$. The CSM DAs were expressed as in Eq.\,\eqref{eq:PDA1}, employing the parameters from Eqs.\,\eqref{eq:paramsCSMDB} and\,\eqref{eq:paramsCSMRL}. Entries marked with an asterisk ($^*$) were inferred using a Gegenbauer polynomial expansion, following the corresponding reference; in those with $(^\dagger)$, we use the simple polynomial form described in Ref.\,\cite{Zhang:2020gaj}. Here $\bar{x}=1-x$.}
\label{tab:momsDA}
\end{table}

\begin{figure}[hptb!]
\includegraphics[width=8.6cm]{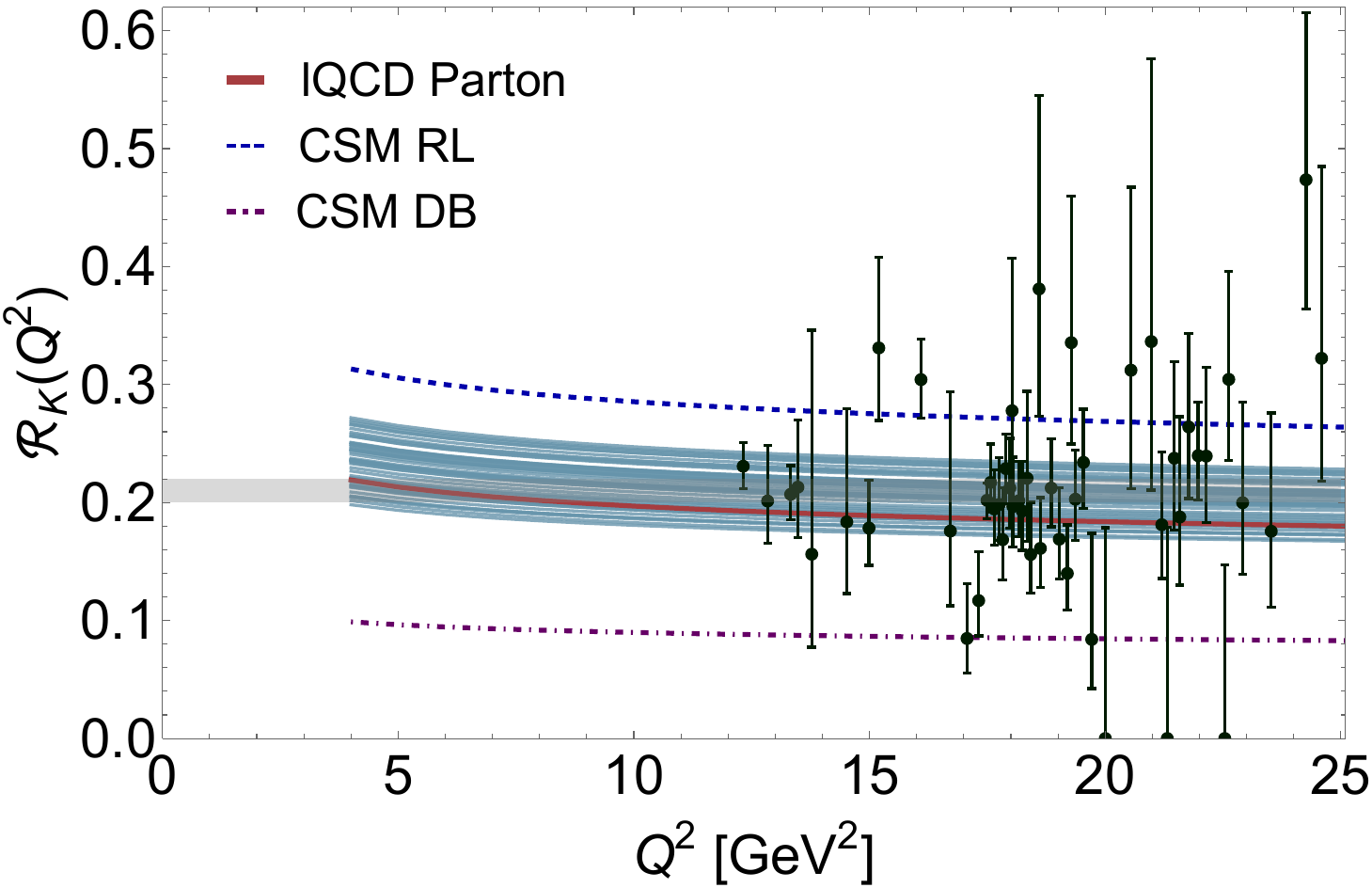}
\caption{Ratio of neutral-to-charged kaon EMFFs, $\mathcal{R}_K$, obtained at LO according to the formulae in Eq.\,\eqref{eq:HSF}-\eqref{eq:RatioDef}, and the DAs discussed in the text. Labels and references are the same as in Fig.\,\ref{fig:PDA}. The experimental data points are those determined by the BESIII collaboration, with the gray rectangle representing the best-fit reported therein, \emph{i.e.} $0.21\pm0.01$~\cite{BESIII:2023zsk}. \label{fig:EFF}}
\end{figure}

\noindent
\emph{Kaon electromagnetic form factors.--} Following the procedure outlined in the previous section, the result of our LO exploration of the ratio of neutral-to-charged kaon EMFFs is presented in Fig.\,\ref{fig:EFF}. The agreement with the experimental data from Ref.\,\cite{BESIII:2023zsk} is evident, to the extent that the fit provided by the BESIII collaboration for this ratio, $0.21\pm0.01$, falls entirely within our set of replicas. In the same figure, we also depict the outcome arising from the lQCD kaon DA -- that determined in Ref.\,\cite{LatticeParton:2022zqc} and depicted herein in Fig.\,\ref{fig:PDA}. The resulting $\mathcal{R}_K$ would also remain within our estimates. For further comparison, Fig.\,\ref{fig:EFF} includes the RL and DB kernel LO expectations for $\mathcal{R}_K$. The former produces the more asymmetric kaon DA among all cases (see Fig.\,\ref{fig:PDA}), leading to a stronger $SU_F(3)$-FSB and, consequently, a larger neutral-to-charged kaon EMFF ratio. In contrast, the DB kernel, which features a broad and slightly asymmetric kaon DA, generates a $\mathcal{R}_K$ with a magnitude three times smaller. Both cases, RL and DB, are outside our acceptable region and, to some extent, could be interpreted as boundaries.

As evidenced by this analysis, employing a LO prescription requires producing a kaon DA with greater splitting between its inverse moments, $\langle 1/x,\,1/\bar{x}\rangle_{\zeta_2}^K$. Only in this way can the correct magnitude for $\mathcal{R}_K$ be obtained. In other words, to be consistent with the available experimental data -- which lies within the $12<Q^2<25\,\text{GeV}^2$ range -- a LO treatment results in a larger skewness for $\varphi_K^l(x;\zeta_2)$. Given that higher-order corrections to the HSF tend to increase $\mathcal{R}_K$,\,\cite{Chen:2024oem}, we expect that incorporating these contributions within our data-driven approach will yield a more symmetric kaon DA. Consequently, the effects of explicit $SU_F(3)$-FSB in shaping the kaon wavefunction would be attenuated, making QCD's mass generation even more dominant. A similar outcome is observed for the kaon distribution function\,\cite{Xu:2024nzp}.


A final piece for examination is the charged kaon-to-pion EMFFs ratio: 
\begin{equation}
\mathcal{R}_{K^+/\pi^+}(Q^2):=F_{K^+}(Q^2)/F_{\pi^+}(Q^2)\,.
\end{equation}
The derivation of the charged kaon EMFF, $F_{K^+}(Q^2)$, follows straightforwardly from our obtained set of replicas and Eq.\,\eqref{eq:HSF}. For the pion, we employ the LO HSF, together with the parametric representation of Eq.\,\eqref{eq:PDA1} for the corresponding DA. We consider two limiting cases, namely: $\langle \xi^2\rangle_{\zeta_2}^\pi=0.24,\,0.28$. The first value lies within the range of lQCD results reported in Refs.\,\cite{RQCD:2019osh,Zhang:2020gaj}, as well as the CSM prediction with the DB kernel\,\cite{Cui:2020tdf,Chang:2013pq}. The second case yields a broader pion DA and aligns more closely with the RL truncation result,\,\cite{Chang:2013pq}, and the lQCD expectation from\,\cite{LatticeParton:2022zqc}. 

\begin{figure}[hptb!]
\includegraphics[width=8.6cm]{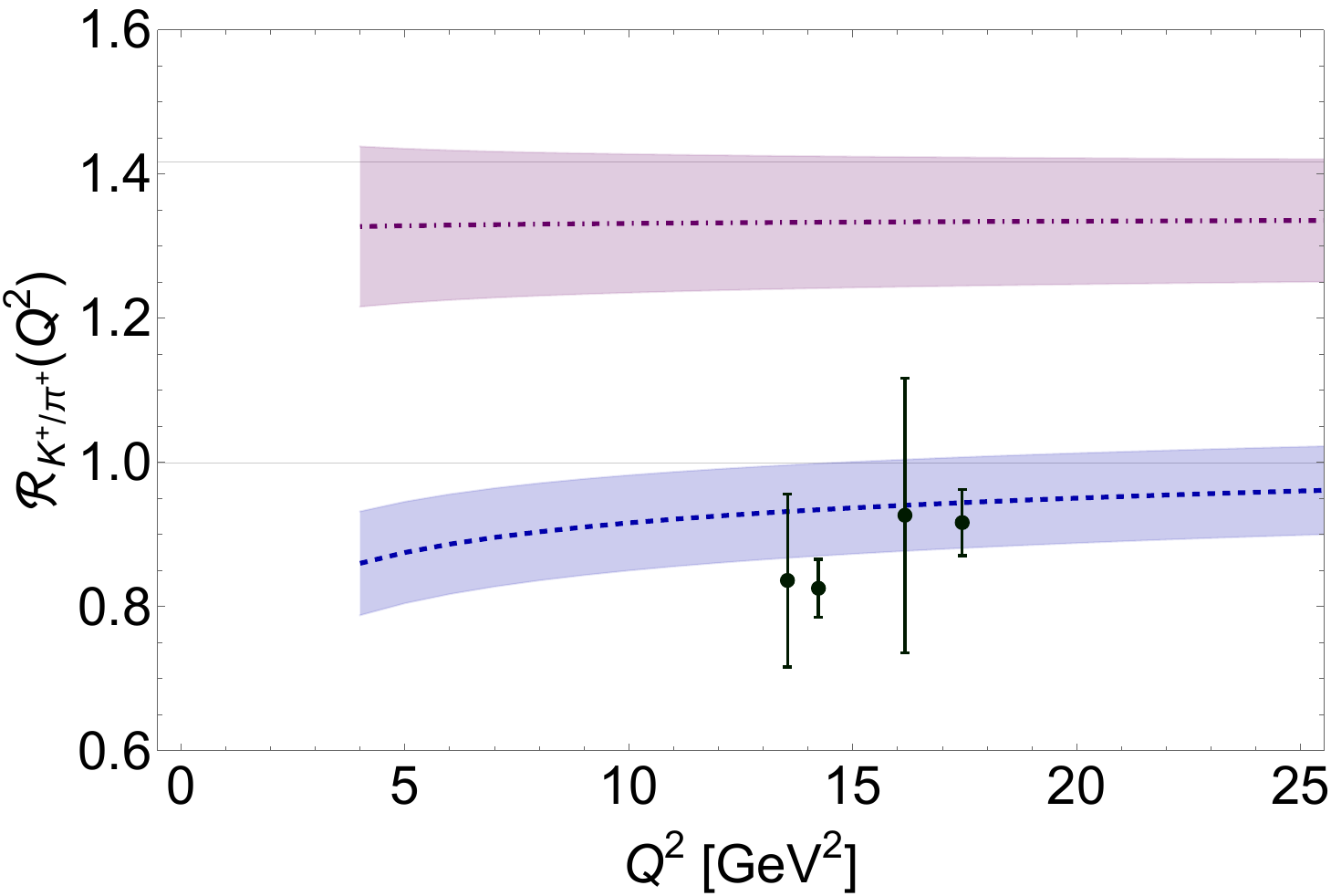}
\caption{Ratio of charged kaon-to-pion EMFFs, $\mathcal{R}_{K^+/\pi^+}$, obtained through the LO HSF, Eqs.\,\eqref{eq:HSF}-\eqref{eq:RatioDef}. For $F_{K^+}(Q^2)$, we adopt the mean of the replica set along with a 1-$\sigma$ error band. For the construction of $F_{\pi^+}(Q^2)$, two DAs are considered: one corresponding to $\langle \xi^2 \rangle_{\zeta_2}^\pi = 0.24$ (purple, dot-dashed), and another to $\langle \xi^2 \rangle_{\zeta_2}^\pi = 0.28$ (blue, dashed). Experimental (timelike) data  taken from Ref.~\cite{Seth:2012nn}. The upper grid line indicates the asymptotic limit, $f_K^2 / f_\pi^2 \approx 1.43$. \label{fig:KtoPi}
}
\end{figure}

The derived result is shown in Fig.\,\ref{fig:KtoPi}. The pion characterized by $\langle \xi^2\rangle_{\zeta_2}^\pi=0.24$ yields a ratio $\mathcal{R}_{K^+/\pi^+}>1$, which rapidly approaches its asymptotic limit $f_K^2 / f_\pi^2$. In contrast, with a broader pion DA, such limit would be reached at a much slower rate. In this case, the LO HSF leads to $\mathcal{R}_{K^+/\pi^+} \lesssim1$ over a wide range of photon virtualities. This profile is more consistent with the timelike experimental data\,\cite{Seth:2012nn}. Nonetheless, since $F_{\textbf{P}}(0)=1$ is fixed by  charge conservation, and the kaon charge radius (a measure of the slope near $Q^2\approx 0$) is smaller than that of the pion\,\cite{Cui:2022fyr,ParticleDataGroup:2024cfk}, the above implies that $F_{\pi^+}(Q^2)$ would intersect $F_{K^+}(Q^2)$ at some point. It is not clear why this would happen. However, tt is important to note that while the dynamics around $Q^2\approx 0$ is governed by vector meson dominance\,\cite{OConnell:1995nse,Gounaris:1968mw}, and the asymptotic behavior by Eq.\,\eqref{eq:HSF}, there is no established prescription for intermediate spacelike values of $Q^2$.

\emph{Summary and scope.--} 
We have introduced a data-driven approach to determine the kaon DA at an experimentally accessible scale of $\zeta=2$ GeV, $\varphi_K^l(x;\zeta_2)$. The procedure described here relies on a simple yet effective two-parameter model for the kaon DA that properly captures the broadening and skewness effects induced by the mechanisms of DCSB and $SU_F(3)$-FSB. The model parameters are determined from recent empirical information on the neutral-to-charged kaon EMFF ratio $\mathcal{R}_K$. For this purpose, the HSFs in QCD are employed at LO level of approximation. The required DAs are fixed in such a way that the model parameters are randomly scanned within a sensible range, informed by lQCD and continuum methods. Our analysis highlights the necessity of employing representations for $\varphi_K^l(x;\zeta_2)$ that  properly reproduce the DA's dilation and skewness, while adhering to the soft endpoint behavior prescribed by QCD. These points are vital in ensuring the correct magnitude for $\mathcal{R}_K$. Moreover, the kaon DAs obtained through our procedure capture the effects of DCSB by yielding  broadened profiles. Conversely, the distributions also exhibit a rather visible asymmetry. Despite this fact, the results of our analysis are fully consistent with well-established findings from continuum and lattice QCD methods. Furthermore, the distortion of the kaon wavefunction is not so marked, suggesting that the explicit breaking of the $SU_F(3)$ flavor symmetry is subdominant, compared to the non-perturbative effects of QCD. In this regard, the incorporation of higher-order effects in the hard-scattering kernels is expected to further reduce the skewness. This shall be addressed elsewhere. On the other hand, the examination of the charged kaon-to-pion EMFF ratio shows that, despite meeting the asymptotic expectations, this value could lie below unity given a sufficiently dilated pion DA. A precise determination of its profile is therefore crucial. Finally, it is anticipated that with future experimental efforts on the pion and kaon EMFFs, the present data-driven approach may prove instrumental in yielding a precise determination of the corresponding wavefunctions.

\emph{Acknowledgments.--} Work supported by: National Natural Science Foundation of China (Grant Nos. 12135007 and 11875169); Spanish Ministry of Science and Innovation (MICINN grant no.\ PID2022-140440NB-C22); and Junta de Andalucía (grant no.\ P18-FR-5057).

\bibliography{main}
\end{document}